\documentclass[namedreferences]{solarphysics}
\usepackage{verbatim}  
\usepackage[hyperref,optionalrh,solaromanenum]{spr-sola-addons} 
\usepackage{graphicx}                    
\usepackage{color}                       
\usepackage{breakurl}                         
\usepackage[flushleft]{threeparttable}
\usepackage{siunitx}
\usepackage{rotating}

\newcommand\araa{{Annu.~Rev.~Astro.~Astrophys.}}%
\newcommand\apj{{Astrophys.~J.}}%
\newcommand\jgr{{J.~Geophys.~Res.}}%
\newcommand\solphys{{Solar~Phys.}}%
\newcommand\ssr{{Space~Sci.~Rev.}}%
\newcommand\apjl{{Astrophys.~J.~Lett.}}%
\newcommand\aap{{Astron.~Astrophys.}}%
\newcommand\aaps{{Astron.~Astrophys.~Supp.}}
\newcommand\apjs{{Astrophys.~J.~Supp.}}%
\begin{document}

\begin{article}

\begin{opening}

\title{Statistical Approach on Differential Emission Measure of Coronal Holes using the CATCH Catalog}

%
\author[addressref={graz},corref,email={stephan.heinemann@hmail.at}]{\inits{S.G.}\fnm{Stephan~G.~}\lnm{Heinemann}\orcid{0000-0002-2655-2108}}
\author[addressref={graz},corref,email={}]{\inits{J.}\fnm{Jonas~}\lnm{Saqri}\orcid{0000-0003-2170-1140}}
\author[addressref={graz,kso},corref,email={}]{\inits{A.M.}\fnm{Astrid~M.~}\lnm{Veronig}\orcid{0000-0003-2073-002X}}
\author[addressref={graz,col},corref,email={}]{\inits{S. J.}\fnm{Stefan~J.~}\lnm{Hofmeister}\orcid{0000-0001-7662-1960}}
\author[addressref={graz},corref,email={}]{\inits{M.}\fnm{Manuela~}\lnm{Temmer}\orcid{0000-0003-4867-7558}}
%
\runningauthor{S.G. Heinemann et al.}
\runningtitle{DEM of Coronal Holes}

\address[id={graz}]{University of Graz, Institute of Physics, Universit{\"a}tsplatz 5, 8010 Graz, Austria }
\address[id={kso}]{Kanzelh{\"o}he Observatory for Solar and Environmental Research, University of Graz, 9521 Treffen, Austria}
\address[id={col}]{Columbia Astrophysics Laboratory, Columbia University, New York, USA}
\begin{abstract}
Coronal holes are large-scale structures in the solar atmosphere that feature a reduced temperature and density in comparison to the surrounding quiet Sun and are usually associated with open magnetic fields. We perform a differential emission measure analysis on the 707 non-polar coronal holes collected in the  Collection of Analysis Tools for Coronal Holes (CATCH) catalog to derive and statistically analyze their plasma properties (i.e. temperature, electron density, and emission measure). We use intensity filtergrams of the six coronal EUV filters from the \textit{Atmospheric Imaging Assembly} onboard of the \textit{Solar Dynamics Observatory}, which cover a temperature range from $ \approx 10^{5.5}$ to $10^{7.5}$\,K. Correcting the data for stray and scattered light, we find that all coronal holes have very similar plasma properties with an average temperature of $0.94 \pm 0.18$ MK, a mean electron density of $(2.4 \pm 0.7) \times 10^{8}$\,cm$^{-3}$, and a mean emission measure of $(2.8 \pm 1.6) \times 10^{26}$\,cm$^{-5}$. The temperature distribution within the coronal hole was found to be largely uniform, whereas the electron density shows a $40\,\%$ linear decrease from the boundary towards the inside of the coronal hole. At distances greater than \SI{20}{\arcsecond} ($\approx 15$\,Mm)  from the nearest coronal hole boundary, the density also becomes statistically uniform. The coronal hole temperature may show a weak solar cycle dependency, but no statistically significant correlation of plasma properties to solar cycle variations could be determined throughout the observed time period between 2010 and 2019.
\end{abstract}

%

\end{opening}


%

\section{Introduction} \label{s:intro}
Coronal holes are large-scale magnetic structures that extend from the solar photosphere into interplanetary space and are characterized by their open-to-interplanetary-space magnetic field configuration. This distinct magnetic topology enables plasma to be accelerated to high speeds of up to $780$\,km\,s$^{-1}$ and escape along the open field lines \citep{schwenn06}. Coronal holes are defined by a lower density and temperature in comparison to the surrounding corona and and are thus observed as large-scale regions of reduced emission in X-ray and extreme ultraviolet (EUV) wavelengths \citep[see reviews by][and references therein]{cranmer2002,cranmer2009}.\\

 Studies using differential emission measure (DEM) techniques on spectroscopic data from the EUV imaging spectrometer onboard Hinode\citep[Hinode/EIS;][]{2011hahn} and the Solar Ultraviolet Measurements of Emitted Radiation on the Solar and Heliospheric Observatory \citep[SOHO/SUMER;][]{2008landi} as well as on SDO/AIA EUV filtergrams \citep{2020saqri} revealed that coronal holes have a peak in the emission at a temperature of $T \approx 0.9$ MK. Coronal Diagnostic Spectrometer (SOHO/CDS) spectroscopy also suggests this \citep{1999Fludra}. This is significantly lower than the temperature of the surrounding quiet corona, which is often estimated at temperatures around $1.4$\,-\,$1.6$\,MK \citep{2008Landi_QS,2011hahn,2013delzanna,2014Mackovjak,2014hahn}. \cite{2018Wendeln} and \cite{2020saqri} used Hinode/EIS and SDO/AIA data to derive DEM profiles and revealed that besides the dominant contribution at around $0.9$ MK a secondary peak around $\approx\,1.4$\,-\,$1.6$ MK was present within the DEM of coronal holes. It was suggested that it is mostly due to the presence of stray light. Electron densities in coronal holes have been estimated ranging from $1.0-2.5 \times 10^{8}$ cm$^{-3}$ \citep{1999Fludra,1999warren,2011hahn,2020saqri}. Using coronagraphic white-light images, \cite{1994Guhathakurta} showed that the density in coronal holes varies in height above the solar surface but not over latitude.\\

The plasma properties (i.e., emission measure, temperature, and density) of coronal holes have been been analyzed using different observations and methods; however, it has been done only in the scope of observational campaigns or case studies but not in the scope of a large statistical approach. By using long-term EUV observations from 2010 to 2019 covering nearly the full Solar Cycle 24, we are able to statistically investigate the distributions of the plasma properties for a large variety of coronal holes of different sizes, and whether these properties change over the solar cycle.\\

In this study, we performed differential emission measure (DEM) analysis using data from the \textit{Atmospheric Imaging Assembly} \citep[AIA:][]{2012lemen_AIA} on-board the \textit{Solar Dynamics Observatory} \citep[SDO:][]{2012pesnell_SDO} on coronal holes of the extensive Collection of Analysis Tools for Coronal Holes (CATCH) catalog\footnote{Vizier Catalog: vizier.u-strasbg.fr/viz-bin/VizieR?-source=J/other/SoPh/294.144} \citep{2019heinemann_catch}. We derive the distribution of plasma properties and their dependence on the solar cycle, and we relate them to the primary coronal hole parameters such as area and magnetic field density.

\section{Methodology}\label{s:methods}

\subsection{Dataset} 
For the presented statistical study we used the CATCH catalog, which contains $707$ observations of well-defined non-polar coronal holes extracted from SDO/AIA $193$\,\AA\ filtergrams. The catalog contains the extracted boundaries and properties of the coronal holes such as area, intensity, signed and unsigned magnetic field strength and magnetic flux including uncertainty estimates. The coronal holes are distributed between latitudes of $\pm 60^{\circ}$ and cover nearly the full Solar Cycle 24 from 2010 to 2019. A description of the catalog has been given by \cite{2019heinemann_catch}. 

\subsection{Data Processing} \label{subs:data}
For the DEM analysis the level 1.6 data (processed with \verb+aia_prep.pro+ and point-spread-function-corrected) of the six coronal channels of AIA/SDO \citep{2012lemen_AIA} was used, namely 94\,\AA\ (Fe \textsc{xviii}), 131\,\AA\ (Fe \textsc{viii}, \textsc{xxi}), 171\,\AA\ (Fe \textsc{ix}), 193\,\AA\ (Fe \textsc{xii}, \textsc{xxiv}), 211\,\AA\ (Fe \textsc{xiv}) and 335\,\AA\ (Fe \textsc{xvi}). To enable us to pre-process all coronal holes in a reasonable amount of time and to enhance the signal-to-noise ratio for the DEM analysis, the data were rebinned by a factor of eight to a plate scale of \SI{4.8}{\arcsecond} per pixel. 

\subsection{Point Spread Function Correction} \label{subs:psf}
From previous DEM studies of coronal holes it is known that stray light and scattered light significantly affects the analysis, since the emission in coronal holes is much lower than the surrounding quiet Sun and active regions \citep{2018Wendeln, 2020saqri}. Therefore, we use the point spread function (PSF) correction available in the SolarSoftware package of the Interactive Data Language (SSW IDL) \textsc{aia psf} (\verb+aia_calc_psf.pro+ written by M.Weber, SAO), to remove contributions from bright sources as well as possible.  \\

\begin{figure} 
 \centerline{\includegraphics[width=.7\textwidth,clip=]{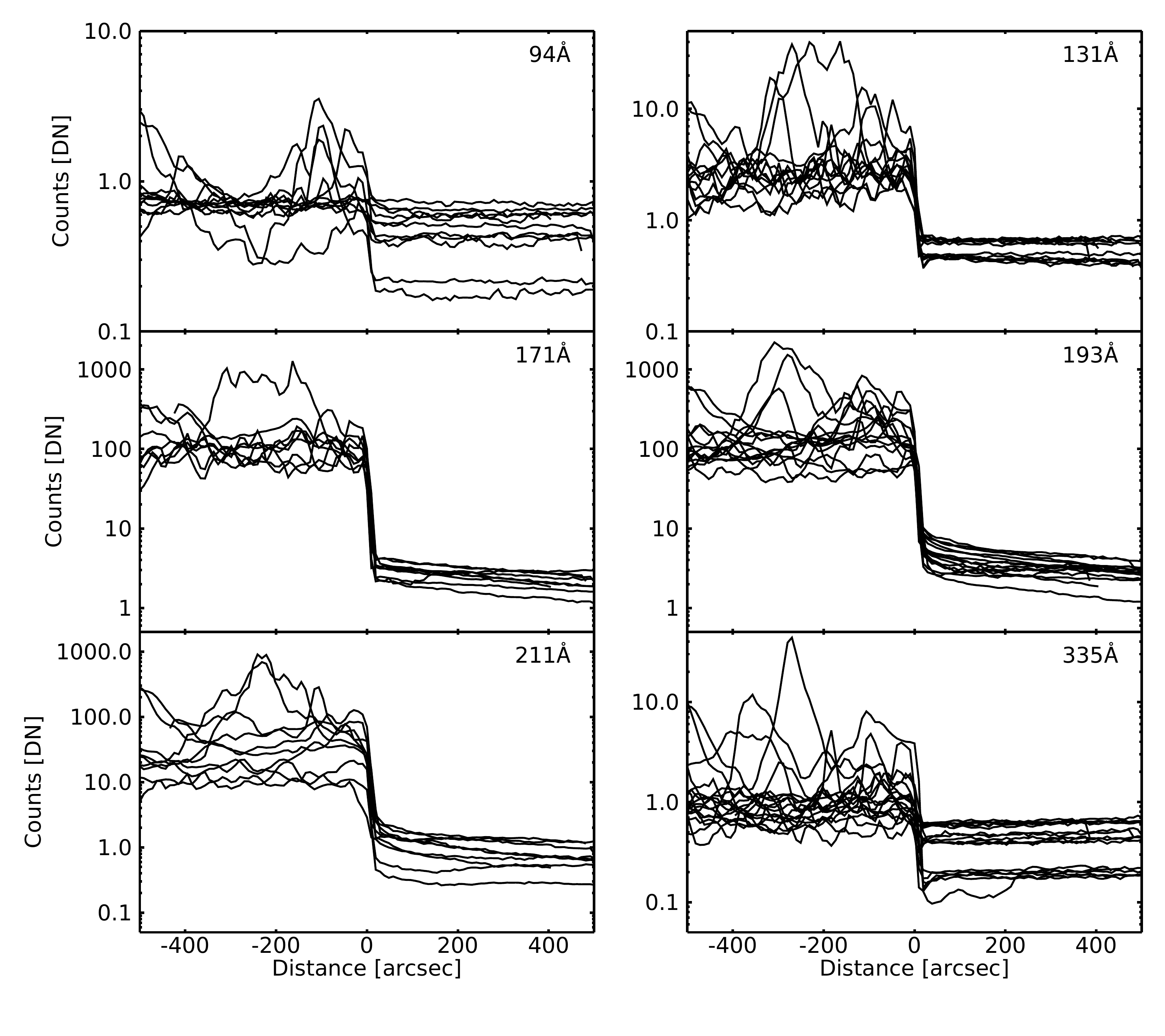}}
 \caption{Superposed intensity profiles across the solar disk during $\approx\,20$ lunar eclipses of level 1.6 data for different wavelengths. The eclipse border is centered on $0$, the negative direction marks the visible solar disk and the positive direction the part covered by the lunar eclipse.}\label{fig:eclipse}
 \end{figure}

We tested the performance of the PSF corrections using lunar-eclipse observations. At the eclipse boundary, the counts should drop to zero and all measured counts are due to stray light and noise. A well-performing PSF correction should show such a behavior. To verify, we analyzed $\approx 20$ lunar eclipses that occurred between 2010 and 2019 in all six wavelengths. Figure~\ref{fig:eclipse} shows for each AIA filter, superposed light profiles of level 1.6 data across the boundaries of multiple solar eclipses. We find that for all channels, some counts remain, however, without an obvious correlation to the solar cycle or the mean intensity of the solar disk. The hot channels (94\,\AA\,, 131\,\AA\,, 335\,\AA ) show no dependence on the distance from the transition, and we assume the remaining counts to be primarily isotropic noise. This seems especially true for the 94\AA\ filter, where the remaining eclipse counts may be in the order of the quiet-Sun counts. The 171\,\AA, 193\,\AA,  and 211\,\AA\ channels display a weak dependence on the distance, which indicates the presence of some uncorrected long-range scattered light \citep{2018Wendeln}. We found no working solution to correct for this; however, we can account for the remaining counts in the DEM analysis by increasing the input error of the counts (DN) in each pixel \citep[see ][]{2020saqri}. For each filter, we derive the remaining counts by averaging over the lunar eclipse light profiles (as shown in Figure~\ref{fig:eclipse}) at a distance between \SI{50}{\arcsecond} to \SI{200}{\arcsecond} from the eclipse boundary. These distances were chosen to avoid contribution of the transition and associated boundary effects in the eclipse data (\SI{50}{\arcsecond}) and because there are no coronal hole pixels with distances exceeding \SI{200}{\arcsecond}. The average eclipse counts for level 1.5 and level 1.6 are listed in Table~\ref{tab:psf}. We find that the level 1.6 data shows strongly reduced remaining eclipse counts in the 171\,\AA, 193\,\AA, and 211\,\AA\ channels.  The other channels are primarily dominated by isotropic noise and comparable between level 1.5 and 1.6 data. The eclipse correction for the DEM input is given as the mean remaining eclipse counts (as stated above and shown in Table~\ref{tab:psf}) plus one sigma. Note, that with this method we ignore the distance dependence and possible effects such as the roughly one-sided illumination of the eclipse in contrast to coronal holes in the center of the disk. This is done because we cannot reliably make an estimate for such effects.

\begin{table}
\caption{Mean lunar eclipse counts (DN) of level 1.5 and level 1.6 data for all wavelengths derived from the light profiles in Figure~\ref{fig:eclipse}, the resulting eclipse correction as well as the mean counts of the 707 coronal holes using the level 1.6 data}\label{tab:psf}
\begin{threeparttable}
\begin{tabular}{lc c c c c c}
  & 94\AA\ & 131\AA\ & 171\AA\ & 193\AA\ & 211\AA\ & 335\AA\  \\ \hline
  level 1.5 & $0.1 \pm 0.1$ & $0.1 \pm 0.1$ & $3.5 \pm 1.0$ & $5.1 \pm 2.1$ & $1.4 \pm 0.8$ & $0.1 \pm 0.1$ \\ 
  level 1.6 & $0.5 \pm 0.2$ & $0.5 \pm 0.1$ & $2.8 \pm 0.6$ & $3.9 \pm 1.3$ & $1.1 \pm 0.5$ & $0.4 \pm 0.2$ \\
eclipse correction & $0.7$ & $0.6$ & $3.4$ & $5.2$ & $1.6$ & $0.6$ \\ \hline
coronal hole & $0.2 \pm 0.3$ & $1.5 \pm 0.8$ & $68.6 \pm 17.2$ & $28.6 \pm 9.6$ & $ 7.3 \pm 3.0$ & $0.4 \pm 0.4$ \\

\hline
\end{tabular}
\end{threeparttable}
\end{table}

\subsection{Differential Emission Measure} \label{subs:dem}
Differential emission measures analysis is the reconstruction of the plasma properties from observed intensities in different wavelengths. The DEM is defined as the emission of optically thin plasma in thermodynamic equilibrium for a specific temperature along the line-of-sight (LoS) and is given by
\begin{equation}
    \mathrm{DEM}(T)=n_{\mathrm{e}}(T)^{2}~\frac{\mathrm{d}h}{\mathrm{d}T},
\end{equation}
with $n_{\mathrm{e}}$ the electron number density as function of the temperature $T$ and $h$ the LoS distance over which the emission observed is integrated \citep[see][ Chapter 4]{1992Mariska}.
According to \cite{2012hannah+kontar}, from the observed intensities of each filter [$I_{\lambda}$] the DEM can be estimated by solving the inverse problem
\begin{equation}\label{eq:2}
    I_{\lambda} = \int_{T}~K_{\lambda}(T)~\mathrm{DEM}(T)~\mathrm{d}T,
\end{equation}
with $K_{\lambda}$ being the instrumental response functions. By following Equation 6 from \cite{2012Cheng} the electron number density ($n_{e}$) for coronal holes can be calculated from the DEM as follows:
\begin{equation}
    n_{e}=\sqrt{\frac{\mathrm{EM}}{h}},
\end{equation}
with the emission measure $\mathrm{EM}=\int \mathrm{DEM}~\mathrm{d}T$ and the LoS integration length [$h$] which we approximate with the hydrostatic scale height $h=\frac{\mathrm{k}_{\mathrm{B}}T}{\mathrm{m}g\mu}$. Due to the open magnetic field that does not vertically constrain the plasma, we assume the hydrostatic scale height to be valid as a first-order approximation. However this is only valid for the regime where the field is mostly vertical and approximately uniform i.e., it is not valid outside of coronal holes. In quiet Sun regions, a height-dependent DEM model using a scale height approximation modeled with an ensemble of multi-hydrostatic loops might be used \citep{2005Aschwanden_book}. We use $g=274$\,ms$^{-2}$, a proton mass of $\mathrm{m}=1.67 \times 10^{-27}$\,kg \citep{2005Aschwanden_book}, $\mu=0.60$ \citep{2009Asplund}, and the median DEM temperature (Equation~\ref{eq:tmed}) for the hydrostatic scale height. The input error of each pixel was calculated using \verb+aia_bp_estimate_error.pro+ considering shot-, dark-, read-, quantum-, compression- and calibration noise to which the eclipse correction is added (see Table~\ref{tab:psf}). \\

To investigate the plasma properties of the coronal holes, we applied a regularized inversion technique developed by \cite{2012hannah+kontar} to reconstruct the DEM from the six optically thin EUV channels of SDO/AIA. As Equation~\ref{eq:2} does not yield a unique solution without further constraints, the code by \cite{2012hannah+kontar} gives the DEM solution with the smallest amount of plasma required to explain the observed emission (zeroth-order constraint). The instrument response function [$K_{\lambda}(T)$] was calculated assuming photospheric abundances (CHIANTI 9 database: \citealt{1997dere_chianti,2019dere_chianti}) and the AIA filter response function available via SSW IDL (\verb+aia_get_response.pro+). We prefer photospheric over coronal abundances as it was shown that the elemental abundances in chromospheric and coronal layers of coronal holes strongly resemble the photospheric ones \citep{1998feldmann,2003feldmann}. For every pixel in each coronal hole we use $60$ equally spaced temperature bins (in $\log_{10}$~space) between $\log_{10}(T) = 5.3$ and $\log_{10}(T) = 6.5$ to calculate the DEM curves. The integral over all bins gives the total emission measure (EM). From the DEM curve the EM-weighted median temperature is calculated as such that:
\begin{equation}\label{eq:tmed}
    \int_{5.3}^{T_{\mathrm{median}}} \mathrm{DEM}~\mathrm{d}T = \frac{\mathrm{EM}}{2}.
\end{equation}
 The median temperature was chosen over a mean or EM-weighted mean temperature because it better describes the asymmetrical DEM profile and thus better represents the dominant emission from coronal holes \citep[also see DEM curves in][]{2020saqri}. For each pixel \textit{i}, we derive the EM-weighted median temperature [$T_{\mathrm{median},i}$], and calculating the mean of all pixels in a coronal hole gives the average coronal hole temperature as $T_{\mathrm{CH}}= \frac{1}{N} \sum T_{\mathrm{median},i}$ with $N$ being the number of coronal hole pixels.

\begin{figure} 
 \centerline{\includegraphics[width=0.7\textwidth,clip=]{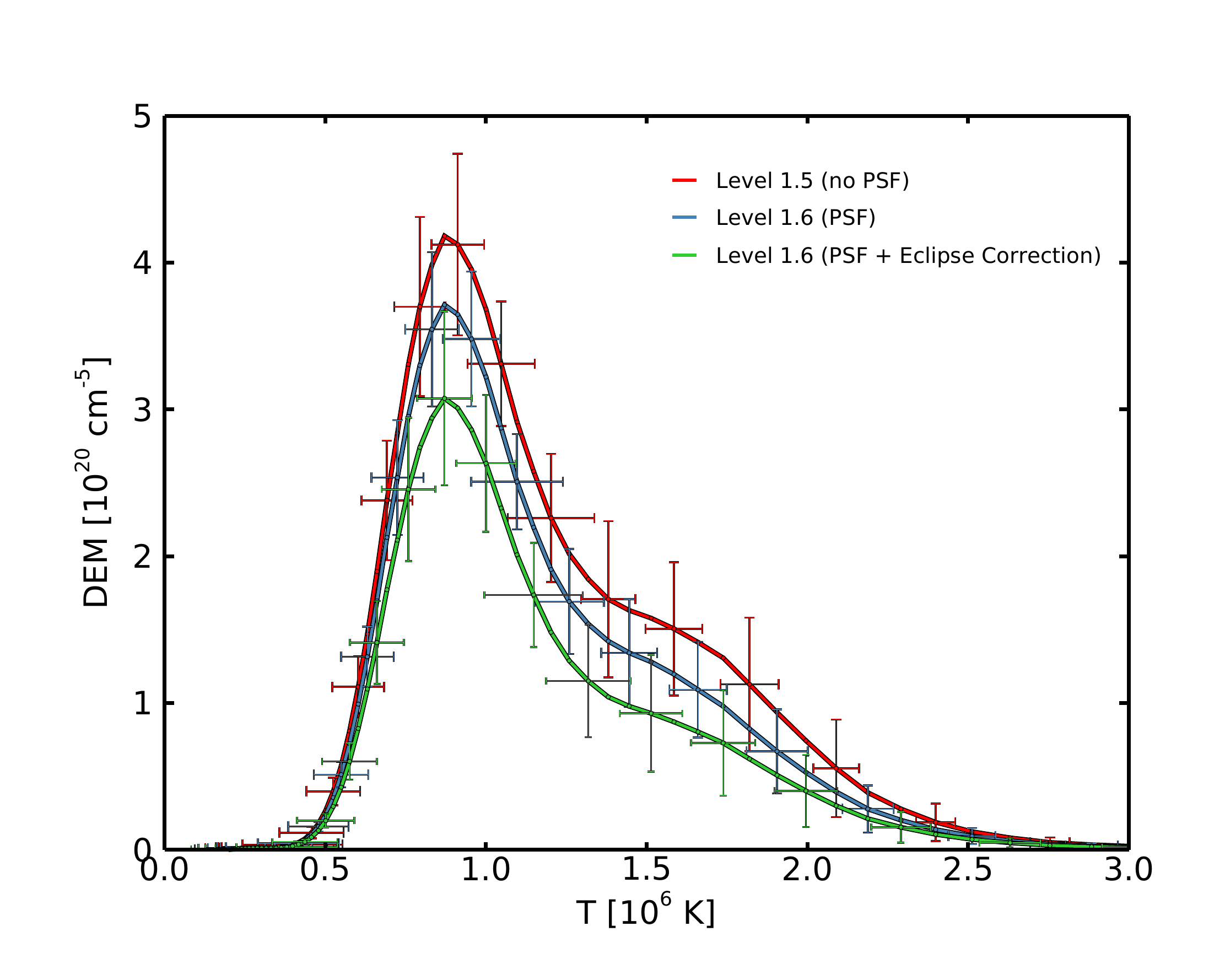}}
 \caption{Average coronal hole DEM for May 29, 2013 using level 1.5 data (red line), level 1.6 (PSF-corrected; blue line), and level 1.6 with eclipse correction (green line). The error bars represent the uncertainties from the DEM calculations of all pixels of the coronal hole. Note that the error bars are shown only for every third bin for better visualization.}\label{fig:dem_comp}
 \end{figure}

In Figure~\ref{fig:dem_comp}, we show the DEM solutions for a coronal hole observed on May 29, 2013 for level 1.5, level 1.6 data and level 1.6 plus applying the eclipse correction. The non-PSF-deconvolved solution shows a peak at high temperatures, which is lower in the PSF-corrected solution and reduces further when considering the remaining counts derived from the Lunar eclipse analysis. This finding supports that the contribution of coronal hole emission at quiet Sun temperatures, also found by \cite{2011hahn} and \cite{2020saqri}, is mainly due to stray light from regions outside the coronal hole.

When calculating the uncertainties, three components have to be considered: the error from the DEM calculation [$\sigma_{\mathrm{\textsc{dem}}}$], which comes from the method and the initial uncertainties in the observations, the variation of the individual pixel values [$\sigma_{\mathrm{\textsc{ch}}}$: standard deviation of pixel values] and the variation of the coronal hole mean values [$\sigma_{\mathrm{\textsc{total}}}$: standard deviation of coronal hole mean values]. The resulting uncertainty can be given as follows:
\begin{equation}\label{eq:err}
    \bar{\sigma} = \sqrt{\bigg(\frac{1}{N_{j}} \sum^{N_{j}}_{j=0} \sqrt{\bar{\sigma}_{\mathrm{\textsc{dem}},j}^2 + \sigma_{\mathrm{\textsc{ch}},j}^2} \bigg)^2 + \sigma_{\mathrm{\textsc{total}}}^2 },
\end{equation}
with $\bar{\sigma}_{\mathrm{\textsc{dem}},j}$ being the mean DEM error and the index $j$ running over all coronal holes.

 \subsection{Correlations} \label{subs:corr}
 The correlation analysis was done using a bootstrapping method \citep{efron1979_bootstrap,efron93_bootstrap} with $> 10^4$ repetitions to derive Pearson correlation coefficients that take into account the uncertainties of the parameters. Additionally, to show that the correlations found do not depend on the data preparation, i.e., on the choice of the PSF nor the estimated correction for the remaining counts, we calculated all correlations with two different PSF deconvolutions (\textsc{aia psf} and a PSF by \citealt{2013Poduval}) and different correction values. This is shown in Tables~\ref{tab:cc_T}\,--\,\ref{tab:cc_em} in the Appendix. Although the absolute values of the plasma parameters can change on average by up to $20\,\%$ depending on the data preparation, the correlations do not change significantly, indicating that the correlations shown in the following sections are reliable estimates.

\section{Results}\label{s:res}
We derived the plasma properties of a set of 707 coronal holes by performing a DEM analysis using the coronal EUV observations by SDO/AIA and obtained the following results. 

\subsection{DEM of Coronal Holes}
\begin{figure} 
 \centerline{\includegraphics[width=1\textwidth,clip=]{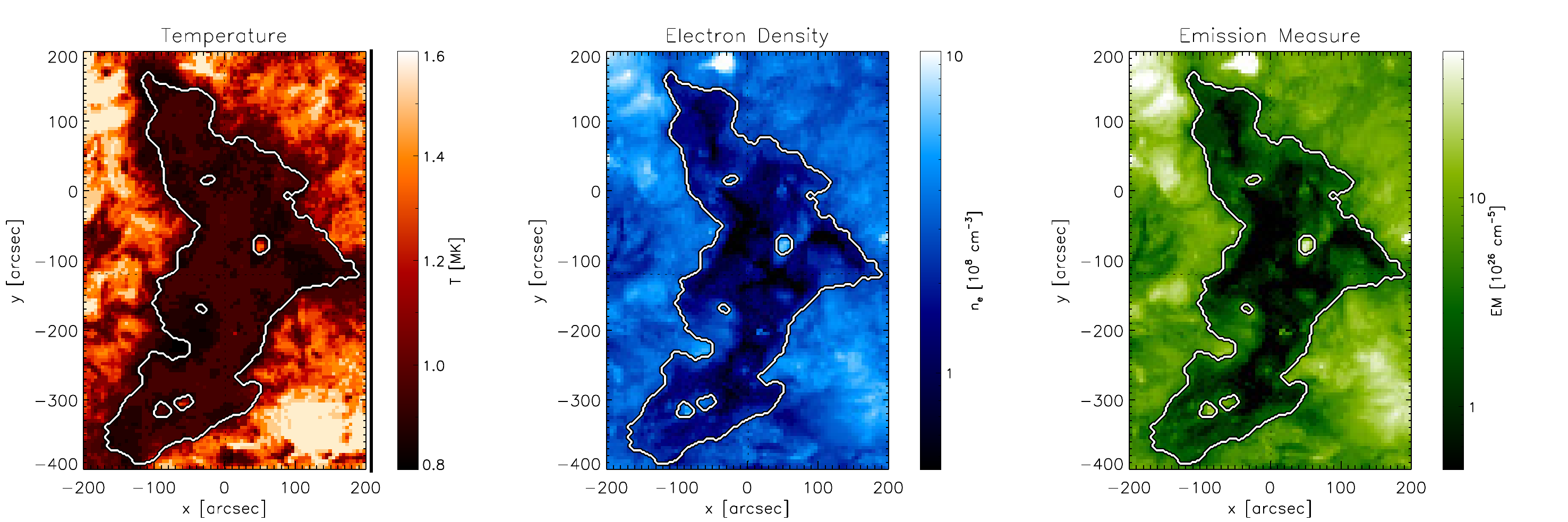}}
 \caption{Example of the DEM analysis of a coronal hole on September 8, 2015. The left panel shows the map of peak temperature, the middle panel the electron density, and the right panel the emission measure. The white contour is the coronal hole boundary as extracted by CATCH \citep{2019heinemann_catch}. The temperature is distributed around a mean of $T_{\mathrm{CH}} = 0.94 \pm 0.05$ MK and thus not well visible. Note that the values outside of the coronal hole boundary should be considered with care as the DEMs were calculated using photospheric abundances, which is applicable in coronal holes but not in the quiet Sun regions, which require coronal abundances.}\label{fig:ch_overview}
 \end{figure}

Figure~\ref{fig:ch_overview} shows an example of the plasma properties (i.e. temperature, density, and emission measure) of a coronal hole and the surrounding quiet Sun areas on September 8, 2015. The temperature shows only small variations around a mean value of $T_{\mathrm{CH}}=0.94 \pm 0.05$\,MK, and is is statistically uniformly distributed inside the coronal hole. The density and emission measure maps show a gradient from the boundary inward with regions of lesser emission and density at some distance from the boundaries. We confirm this statistically in Section~\ref{subsec:d2b}.\\

\begin{figure} 
 \centerline{\includegraphics[width=0.8\textwidth,clip=]{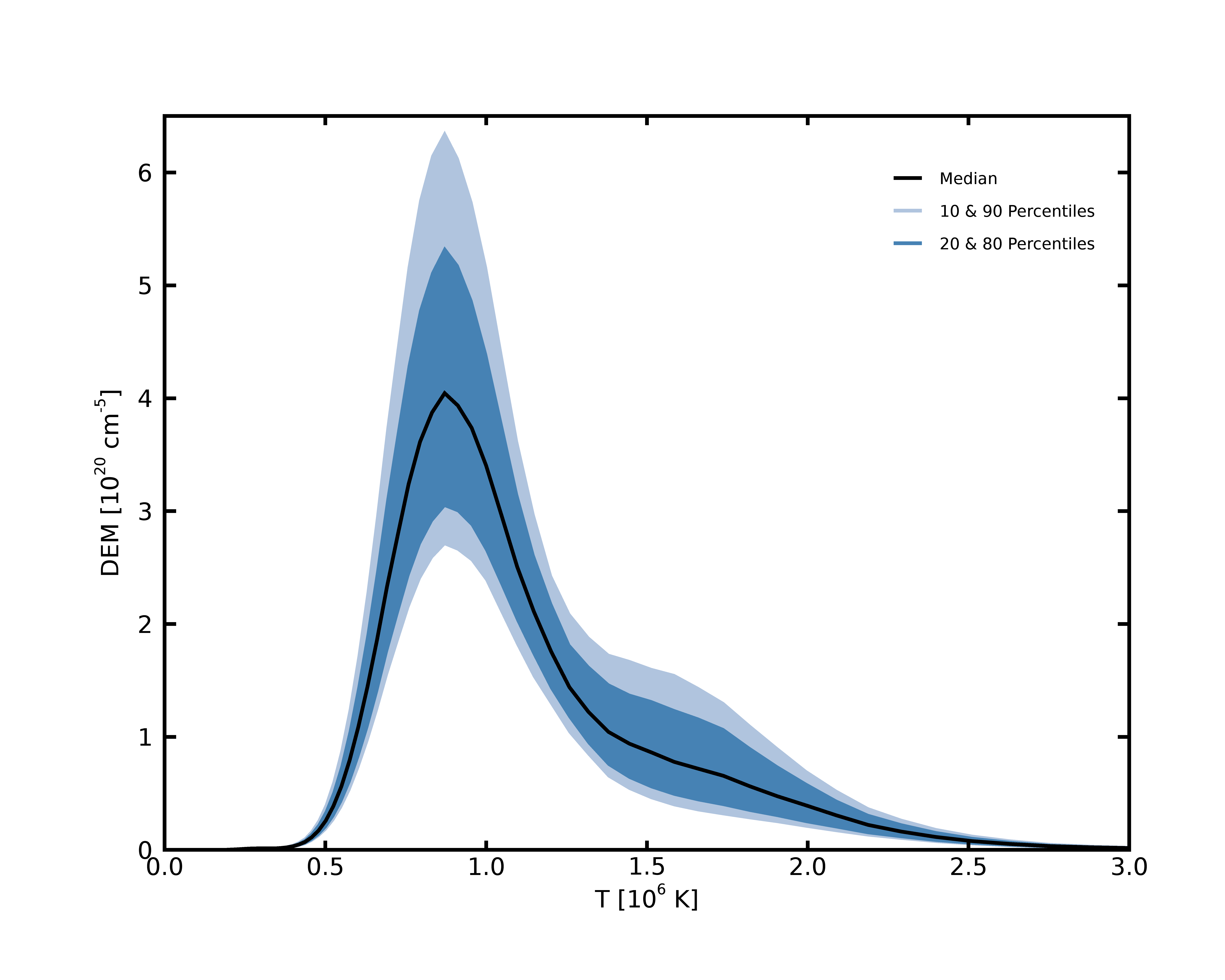}}
 \caption{Median, 80th and 90th percentiles of 707 superposed mean coronal holes DEM curves. The black line gives the median, the blue shaded area the 20th and 80th percentiles, and the gray shaded area the 10th and 90th percentiles. }\label{fig:dem_superposed}
 \end{figure}

 Figure~\ref{fig:dem_superposed} shows the median and the 80th and 90th percentiles of the superposed mean DEM curves of all coronal holes to present the range of values found (note that the errors in the DEM are not shown for this figure). We found that the DEM of a coronal hole is gaussian shaped around a peak temperature of roughly $0.9$\,MK with a tail towards the higher temperatures. We note that the curve is not meaningful for temperatures lower than $0.5$\,MK because the DEM is not well-constrained by the AIA filters \citep[see][for filter sensitivity and response curves]{2012lemen_AIA}. The superposed DEM curve shows that the shape is very similar for all coronal holes, the major variation is in the height of the peak which can vary up to a factor of two. A contribution at quiet Sun temperatures ($\approx~1.4$\,--\,$1.6$\,MK), which is often interpreted as stray light, is hinted at.\\

\begin{figure} 
 \centerline{\includegraphics[width=0.8\textwidth,clip=]{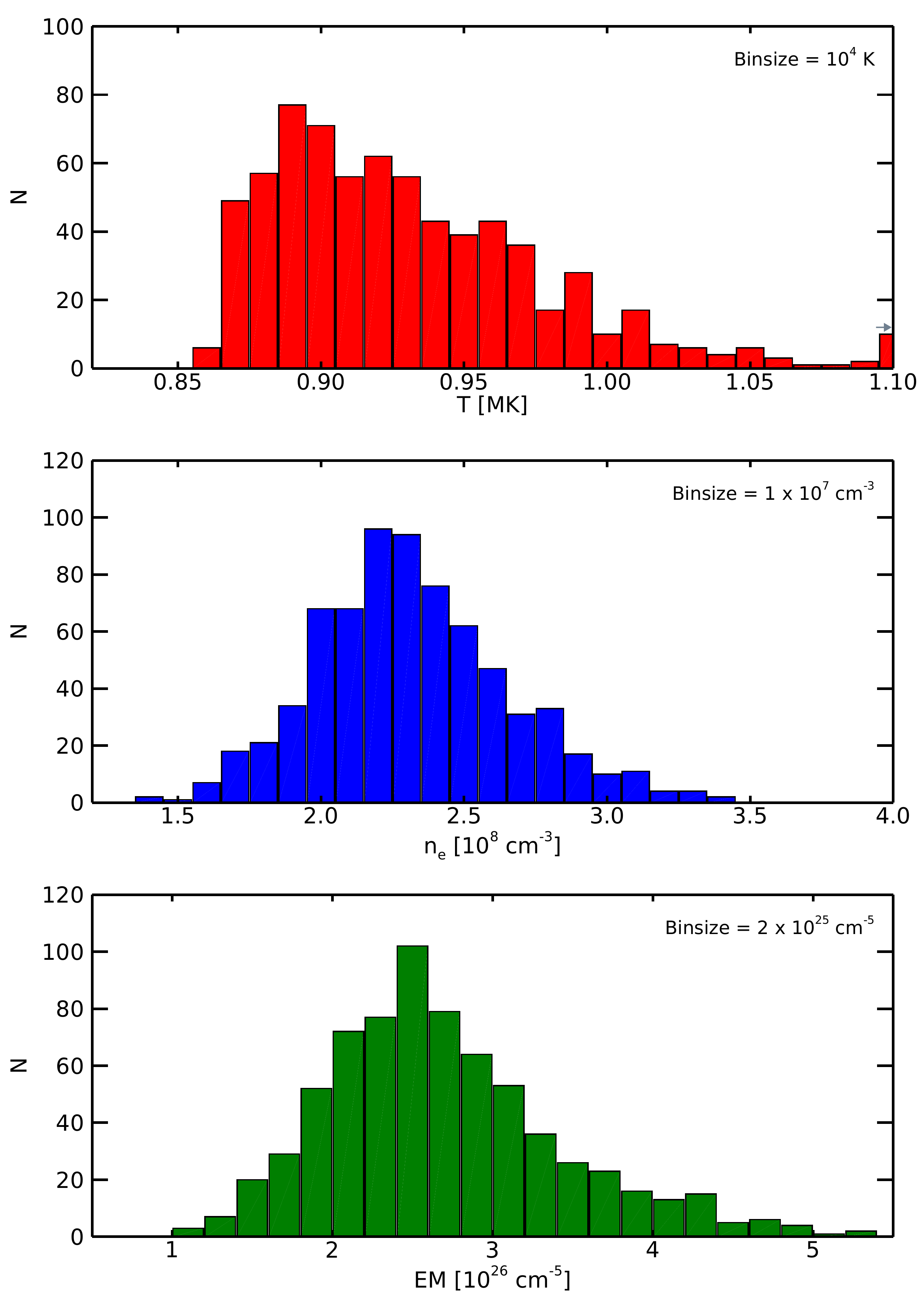}}
 \caption{Distribution of the average coronal hole plasma properties derived from the entire sample of 707 coronal holes. From top to bottom: the temperature, the electron density, and the emission measure. In the temperature histogram we use an overflow bin for all values $>1.1$\,MK. For better visualization the uncertainties have not been included in the figure, but in the calculations of the correlation coefficients they have been considered. }\label{fig:dem_hist}
 \end{figure}

From the average DEM curve of each coronal hole, we derived the temperature, electron density, and emission measure according to the equations in Section~\ref{subs:dem}. Figure~\ref{fig:dem_hist} shows the distribution of the average coronal hole plasma properties. By averaging over the 707 coronal holes, we find an average coronal hole temperature [$T$] of $0.94 \pm 0.18$\,MK, with a derived minimum of $0.86$\,MK and a maximum of $1.32$\,MK. For all coronal holes under study, the average temperature [$T_{\mathrm{CH}}$] is significantly lower than average quiet Sun temperatures. The electron density is distributed around $(2.37 \pm 0.72) \times 10^{8}$\,cm$^{-3}$, with a minimum of  $1.43 \times 10^{8}$\,cm$^{-3}$ and a maximum of $3.47 \times 10^{8}$\,cm$^{-3}$. The emission measure was found in a range between $1.15 \times 10^{26}$\,cm$^{-5}$ and $5.76 \times 10^{26}$\,cm$^{-5}$ with a mean of $(2.79 \pm 1.57) \times 10^{26}$\,cm$^{-5}$.

\subsection{Plasma Properties over the Solar Cycle}
\begin{figure} 
 \centerline{\includegraphics[width=1\textwidth,clip=]{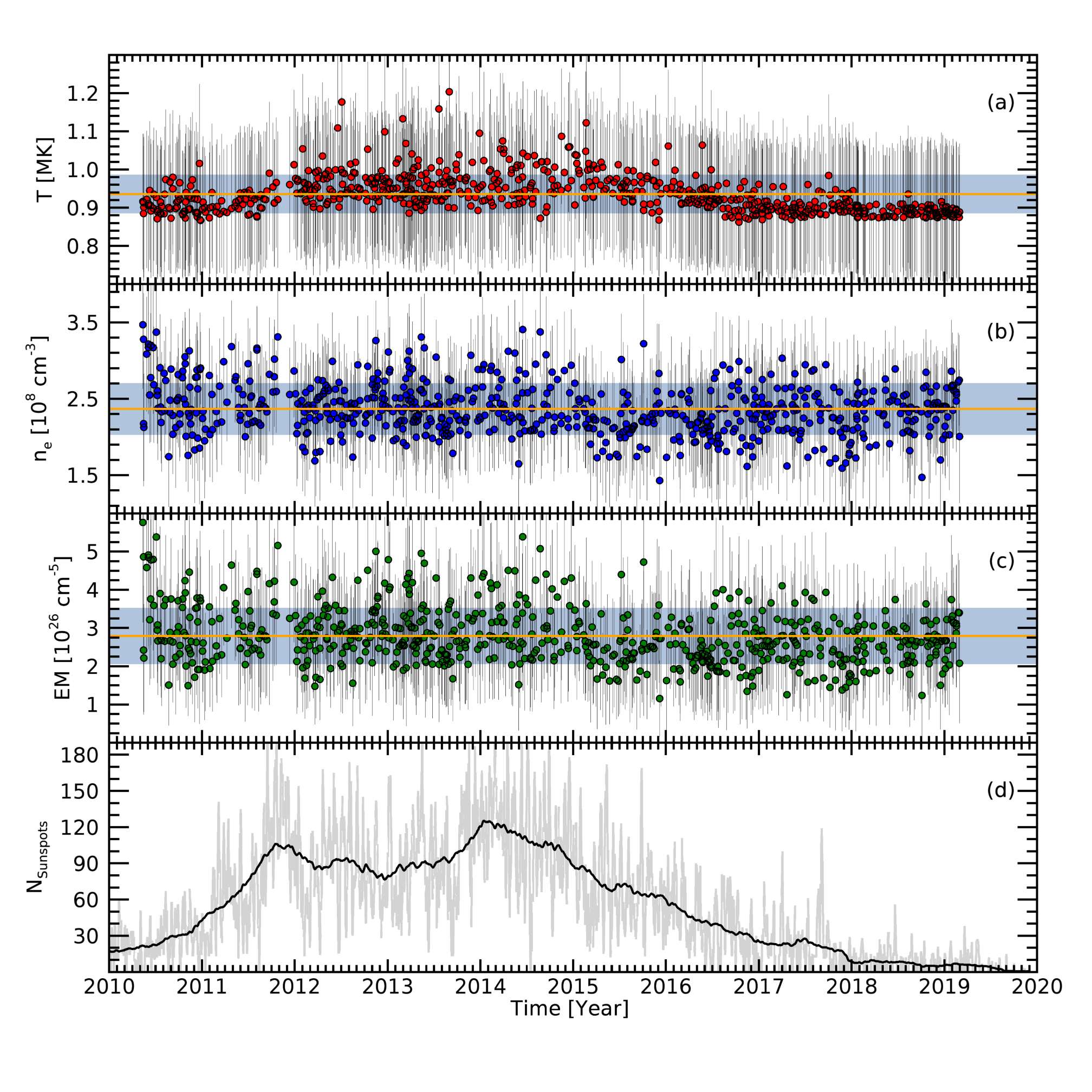}}
 \caption{ Evolution of the coronal hole plasma properties as a function of time. Panel a shows the average coronal hole temperature, panel b the mean electron density, and panel c the mean emission measure of the coronal holes. The orange lines represent the mean of the average values, the shaded area represents the $1\sigma$ uncertainty. The vertical bars represent the uncertainties of each coronal hole ($=\sqrt{\bar{\sigma}_{\mathrm{\textsc{dem}}}^2 + \sigma_{\mathrm{\textsc{ch}}}^2}$). In panel d the daily sunspot number (gray line) and smoothed daily sunspot number (black line) as provided by the SIDC/SILSO are shown.}\label{fig:dem_cycle}
 \end{figure}

Due to the large number of coronal hole observations in the CATCH catalog that span from 2010 to 2019, we can study the plasma properties over almost the full Solar Cycle 24 and investigate a possible dependence on solar activity. In Figure~\ref{fig:dem_cycle} we present the calculated plasma properties together with the international sunspot number (bottom panel) as functions of time. The coronal hole temperature shows small variations over time, which seem to follow the solar activity cycle. When considering the uncertainties of the average coronal hole plasma properties, we find only a very weak correlation to solar activity as approximated by the smoothed sunspot number provided by WDC-SILSO\footnote{Royal Observatory of Belgium, Brussels: www.sidc.be/silso/} with a Pearson correlation coefficient of $cc_{\mathrm{T}}=0.13$ within a $90\,\%$ confidence interval [CI$_{90\%}$] of $[0.07,0.19]$. However, we can quantify that the spread (standard deviation of the average coronal hole values) during the maximum of the solar cycle ($2012-2015$) is twice the spread than in the minimum ($2016-2019$) with $\sigma_{\mathrm{T,max}}=0.06$\,MK and $\sigma_{\mathrm{T,min}}=0.03$. It is worth mentioning that when neglecting the uncertainties and only considering the average values of the plasma parameters derived, a fair correlation of the temperature to the average sunspot number can be found ($cc_{\mathrm{T}} = 0.57$, CI$_{90\%} = [0.52,0.61]$). This difference in the correlation coefficients between the average values with and without the uncertainties considered is attributed to the fact that the uncertainties are larger than the variations of the average values. Neither the emission measure nor the electron density show any clear solar cycle dependence ($cc_{\mathrm{n}_{\mathrm{e}}}=0.04$, CI$_{90\%} =[-0.02,0.10]$  and $cc_{\mathrm{\textsc{em}}}=0.07$, CI$_{90\%} =[0.01,0.14]$).

\subsection{Correlation of Coronal Hole Properties}
\begin{figure} 
 \centerline{\includegraphics[width=0.8\textwidth,clip=]{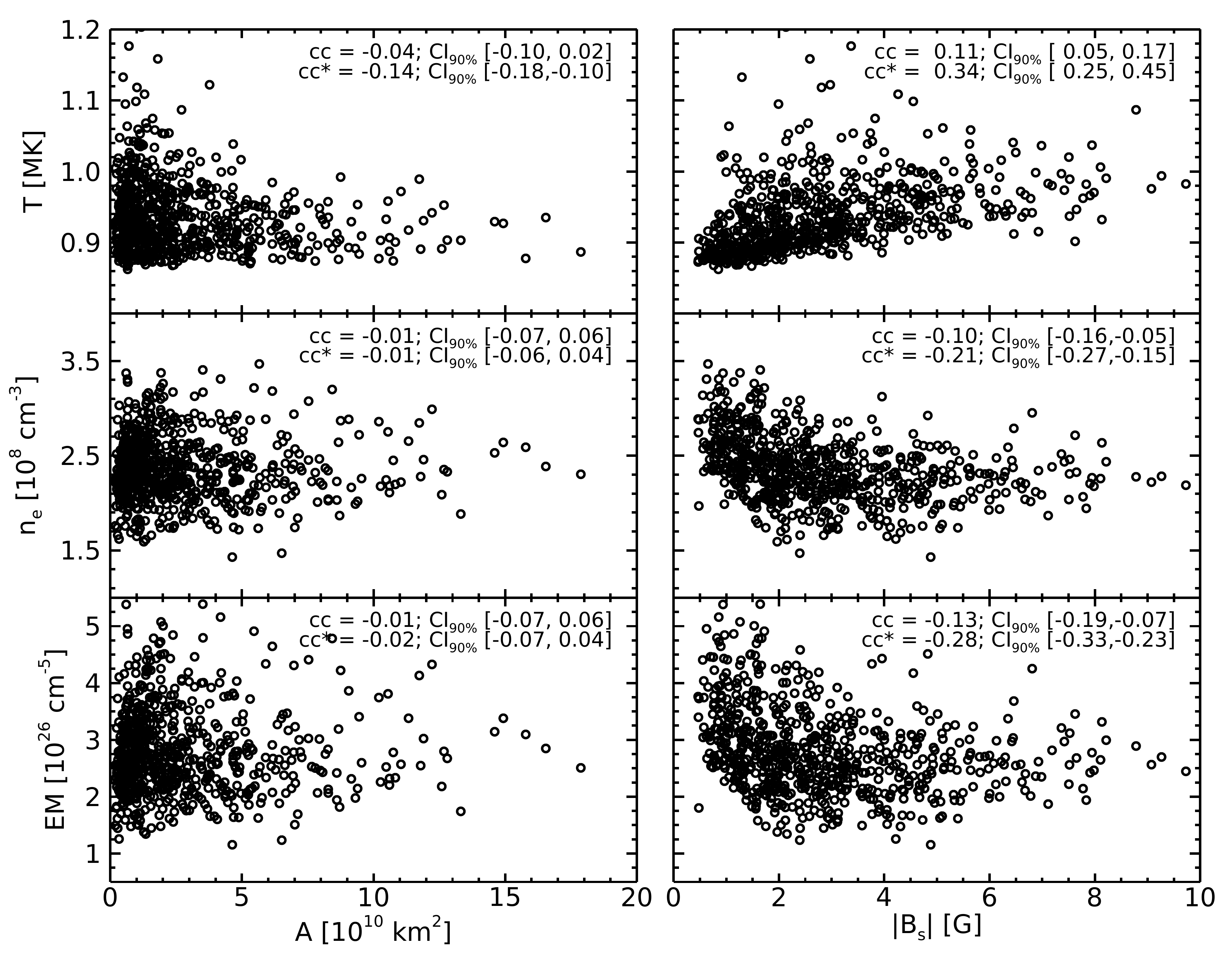}}
 \caption{ Scatter plots of average coronal hole temperature, density and emission measure against area (left) and the mean of the signed magnetic field density (right) from the CATCH catalog \citep{2019heinemann_catch}. The Pearson correlation coefficients claculated with (cc) and without uncertainties (cc*) are given in each respective panel.}\label{fig:dem_areamag}
 \end{figure}

To investigate how plasma properties are correlated to morphological and magnetic coronal hole properties, we plot the average coronal hole temperature [$T_{\mathrm{CH}}$], electron density [$n_{\mathrm{e}}$] and emission measure [$EM$] as function of coronal hole area and signed mean magnetic field density, which were obtained from the CATCH catalog. This is shown in Figure~\ref{fig:dem_areamag}. The left column shows the plasma properties plotted against the coronal hole area and the right column against the magnetic field density. The Pearson correlation coefficients for the plasma properties against the coronal hole area reveal no correlation, with $cc_{\mathrm{T}} = -0.04$, CI$_{90\%} =[-0.10,0.02]$; $cc_{n_{\mathrm{e}}} = -0.01$, CI$_{90\%} =[-0.07,0.06]$; and $cc_{\mathrm{\textsc{em}}} = -0.01$, CI$_{90\%} =[-0.07,0.06]$ for temperature, electron density, and emission measure respectively. With a Pearson correlation coefficient of $cc_{n_{\mathrm{e}}} = -0.10$, CI$_{90\%} =[-0.16,-0.04]$ and $cc_{\mathrm{\textsc{em}}} = -0.13$, CI$_{90\%} =[-0.19,-0.07]$ the density and emission measure show no clear correlation to the mean magnetic field density. The average coronal hole temperature shows a weak trend in the average values when neglecting the uncertainties ($cc_{\mathrm{T}} = 0.34$, CI$_{90\%} =[0.25,0.45]$). However, when including the uncertainties of the derived values in the calculation of the correlation coefficient (according to Section~\ref{subs:corr}) we cannot statistically quantify the correlation ($cc_{\mathrm{T}} = 0.11$, CI$_{90\%} =[0.05,0.17]$).

\subsection{Spatial Distribution of Plasma Properties in Coronal Holes} \label{subsec:d2b}
\begin{figure} 
 \centerline{\includegraphics[width=0.7\textwidth,clip=]{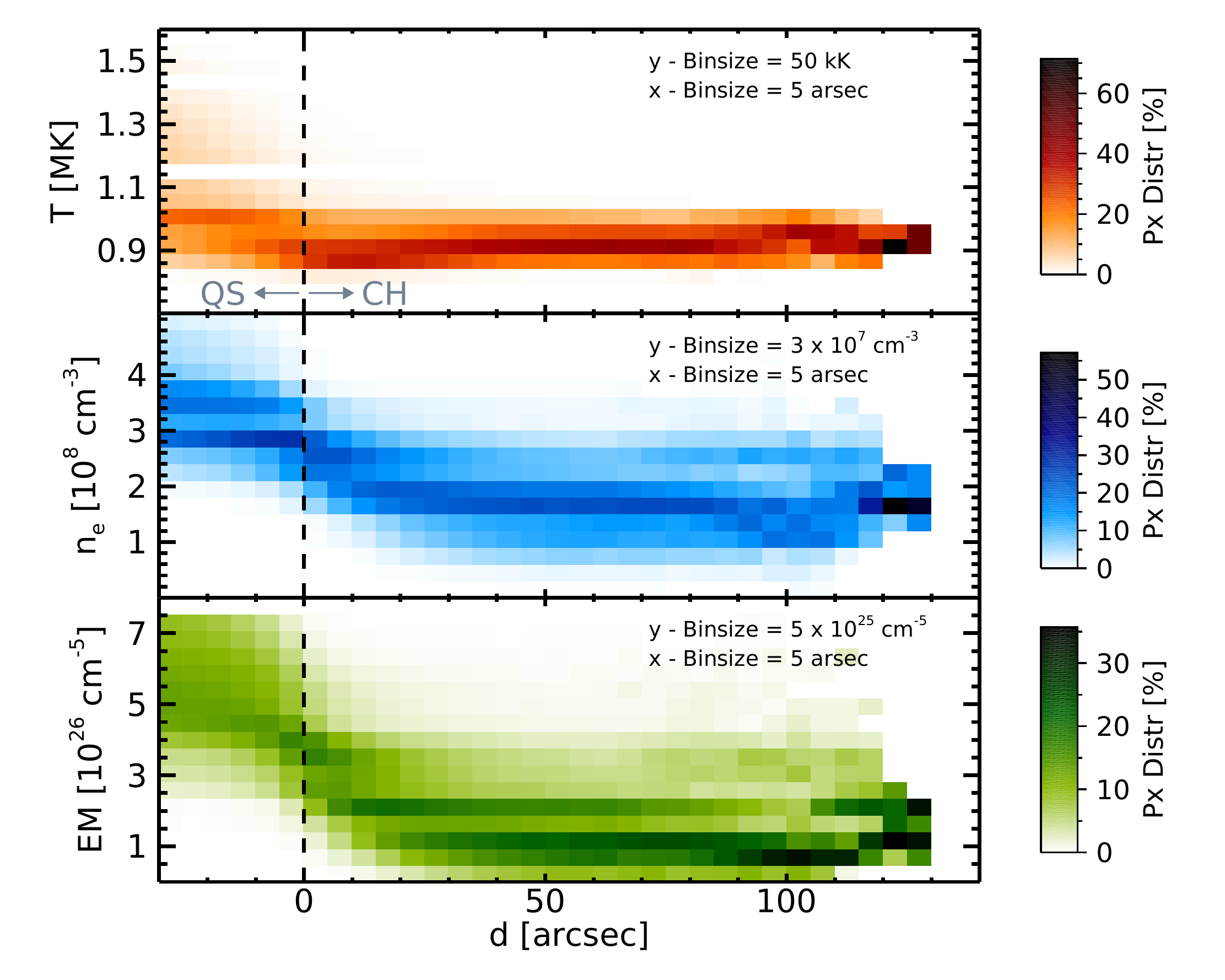}}
 \caption{Distribution of the plasma properties of individual pixels as a function of the distance to the nearest coronal hole boundary [$d$]. For each vertical bin (of a size of \SI{5}{\arcsecond}) the normalized distribution of pixels and their temperature, density, and emission measure (from top to bottom) value was calculated. A darker shade represents a higher percentage of pixels in the according bin. Note that the distributions with $d > $\SI{100}{\arcsecond} are very uncertain because of the low amount of pixels available. Thus, the focus is on the region between $0$ to $\approx $\SI{80}{\arcsecond}. Such a large minimum distance to the coronal hole boundary is only possible in very large coronal holes, which are rare. The negative distances represent pixel outside of the coronal hole, which should be considered with care as discussed in Section~\ref{subsec:d2b}}\label{fig:d2b}
 \end{figure}

In addition to the average coronal hole plasma properties, we investigated how the properties are spatially distributed within the coronal hole. To this aim, we calculated for each coronal hole pixel the distance to the closest coronal hole boundary [$d$ in arcsec] and derived the temperature, electron density, and emission measure of each individual pixel as function of the distance, which is shown in Figure~\ref{fig:d2b}. In each vertical bin of a size of \SI{5}{\arcsecond}, the pixel distribution of $T_{\mathrm{median}}$, $n_{\mathrm{e}}$, and EM is given. Each bin is normalized to reflect the probability for a pixel in a given distance bin to have a certain value. Additionally to the pixels within the coronal holes, the DEM for pixels surrounding the coronal hole boundary was calculated (represented by negative distances) to show the general trend outside of the coronal hole.
We note that the calculations outside are not reliable as they were calculated in the same way as for coronal hole pixels, but here the assumptions of a hydrostatic scale height and photospheric abundances are not valid. Figure~\ref{fig:d2b} shows that the temperature (top panel) within the coronal holes is very uniform and does not depend on the distance from the EUV extracted boundary. Near the boundary, small variations are seen (within $\approx$\,\SI{20}{\arcsecond}) and the temperature does not change strongly at the coronal hole boundary. For the electron density as well as for the emission measure we find a dependence on the distance to the closest coronal hole boundary. Within a distance of \SI{20}{\arcsecond} from the coronal hole boundary, on average the electron density drops by a factor of $\approx\,2$ from roughly $2.8$ to $1.5 \times 10^{8}$\,cm$^{-3}$ and the emission measure by a factor of $\sim 2.5$ from $4.0$ to $1.5 \times 10^{8}$\,cm$^{-5}$. At distances $>$\,\SI{20}{\arcsecond} the dependence ceases and the individual pixel are distributed around $(1.87 \pm 0.61) \times 10^{8}$\,cm$^{-3}$ for the density and around $(2.13 \pm 1.32) \times 10^{26}$\,cm$^{-5}$ for the emission measure (also considering the individual DEM errors). These values were derived for all pixels that are located at distances larger than \SI{20}{\arcsecond} from the closest coronal hole boundary inside coronal holes. When comparing the average pixel values for the electron density and emission measure at distances smaller and larger than \SI{20}{\arcsecond} from the closest coronal hole boundary, we find that the mean pixel densities close ($d < $ \SI{20}{\arcsecond}) to the boundary are a factor $1.3$ higher than further away ($d > $\SI{20}{\arcsecond}). For the emission measure we find a factor of $1.5$. The gradient in the electron density and emission measure shows that the coronal hole boundary extracted using an intensity threshold technique is a good tracer for an area of reduced 
density.

\section{Discussion}\label{s:disc}

The coronal hole properties derived from the presented statistical DEM study are in good agreement with the results of individual case studies and observation campaigns. \cite{1999Fludra} used the CDS on SOHO to derive coronal hole densities and temperatures as a function of height. The temperatures range from $0.75$\,MK at a radial distance of $1.0$\,R$_{\astrosun}$ to $0.85$\,MK at a radial distance of $1.1$\,R$_{\astrosun}$. This approximately agrees with the results derived in this study ($T = 0.94 \pm 0.18$\,MK). The derived average coronal hole temperatures are also in good agreement with the $0.9$\,MK derived by \cite{2008landi}, \cite{2011hahn}, \cite{2018Wendeln} and \cite{2020saqri}. Further, it is notable that the temperature is almost uniformly distributed  ($\sigma_{\mathrm{T}} = 0.05$\,MK) within the coronal holes. \\

When using the DEM analysis to derive electron densities, multiple assumptions have to be made. The densities are calculated over the emission of a LoS column, which we defined as the hydrostatic scale height, which is believed to be a reasonable first-order approximation due to the open field lines in coronal holes. In the quiet Sun a hydrostatic scale height cannot be used due to the abundance of primarily closed fields. As the magnetic topology at the coronal hole boundaries is not well known, it is unclear how valid our assumptions are in this regime. Additionally, it is known that stray light in EUV images can contribute up to $40\,\%$ of the derived electron density value \citep{2012Shearer}, and it is unclear, even after a correction, how much stray and scattered light still remains in the images \citep{2013Poduval}. The derived electron densities of $(2.36 \pm 0.72) \times 10^{8}$\,cm$^{-3}$ are in fair agreement with the DEM study of the evolution of one particular coronal hole performed by \cite{2020saqri}, who derived values between $1.9$ and $2.4  \times 10^{8}$\,cm$^{-3}$, and \cite{1999warren}, who derived values between $1.8$ and $2.4  \times 10^{8}$\,cm$^{-3}$. \cite{2011hahn} used Hinode/EIS spectroscopy of a polar coronal hole and derived density values of  $1.0  \times 10^{8}$\,cm$^{-3}$ and  $1.8  \times 10^{8}$\,cm$^{-3}$ from the Fe \textsc{viii} and Fe \textsc{xiii} line ratios, respectively. They suggested that the density derived from the cooler lines (Fe \textsc{viii}) probes the cool coronal hole plasma and the hotter lines show a contribution from quiet Sun plasma. This also agrees with the results by \cite{2019pascoe} that estimated coronal hole electron densities to be around $\sim\,10^{8}$ cm$^{-3}$. Figure~\ref{fig:d2b} shows that the density decreases as a function of the distance from the coronal hole boundary up to $d \approx $ \SI{20}{\arcsecond}. Under the assumption that the PSF correction used does indeed remove most of the stray light, this result suggest that the extracted coronal hole boundary does not represent a vertical separation of two magnetic regimes but rather the position where the emission drop is the strongest. Thus, the LoS column might represent a mixture of coronal hole plasma emission and quiet Sun plasma emission, maybe due to field lines that are bent rather than radial (e.g. forming an inclined separation between coronal hole and quiet Sun). We notice that the decrease in electron density from the coronal hole boundary to $d\,\approx\,$\SI{20}{\arcsecond} within the coronal hole is approximately $30\,\%$, which is smaller than the $50\,\%$ that has been found by \cite{1997Doschek}.\\ 

The plasma properties show no correlation with the area and magnetic field density of coronal holes. This suggests that the size of a coronal hole does not determine the plasma properties nor vice versa (we find this to be valid for non-polar coronal holes). A trend can be seen in the correlation of the mean magnetic field density and the average coronal hole temperature, however this is not significant ($cc_{\mathrm{T}} = 0.11$, CI$_{90\%} =[0.05,0.17]$) when considering the large uncertainties.\\

The investigation of the plasma properties over the course of the solar cycle between 2010 and 2019 revealed only small variations in density and emission measure. This is especially intriguing as other coronal hole parameters indeed reveal a solar cycle dependence. \cite{2019heinemann_catch} found that the mean $193$\AA\ intensity and the mean magnetic field density show a dependence on solar activity and so does the long-term evolution of large coronal holes \citep{2020Heinemann_chevo}. Only the average coronal hole temperature does show some variation, which might be linked to the solar cycle where slightly higher spreads during solar maximum are observed than during solar minimum. However, the Pearson correlation coefficient does only show a very weak correlation of $cc_{\mathrm{T}}=0.13$ and CI$_{90\%} = [0.07,0.19]$. \cite{2010Doyle} showed that the temperature in coronal holes varies with the solar cycle between $0.8$\,MK in solar minimum and $1.04$ in solar maximum, which we find to be in fair agreement with this study when considering the average values in the correlation, without taking into account the uncertainties ($cc_{\mathrm{T}} = 0.57$, CI$_{90\%} = [0.52,0.61]$). \\

\section{Summary and Conclusions}\label{s:sum}
In this statistical study on the DEM analysis of coronal holes we investigated temperature, electron density, and emission measure of the 707 coronal holes from the CATCH catalog and analyzed their distribution, variability over Solar Cycle 24, and correlations to the coronal hole area and magnetic field density. Our major findings can be summarized as follows.

\begin{enumerate}
    \item \textit{DEM:}~~The shape of the average DEM curve for coronal holes is very stable and resembles a Gaussian profile centered around a peak temperature with an extended tail towards higher temperatures.
    
    \item \textit{Temperature:}~~We find that coronal holes show a EM-weighted median temperature of $0.94 \pm 0.18$\,MK, which is in accordance with previous studies. Additionally, we found that the temperature is spatially very uniform inside the coronal hole.
    
    \item \textit{Electron Density:}~~Using the DEM analysis we derive electron number densities for the coronal holes to be $(2.4 \pm 0.7) \times 10^{8}$\,cm$^{-3}$. The density profile within coronal holes decreases from the boundary to \SI{20}{\arcsecond} by $\approx 30\,\%$ and approaches a constant level further inside. 
    
    \item \textit{Solar Activity:}~~We observe only small variations in the temperature, electron number density, and emission measure during the period from 2010 to 2019 but these appear not to be correlated to changes in solar activity. Although the average coronal hole temperature may hint toward some solar cycle dependence, due to the uncertainties this dependence is not significant.
        
\end{enumerate}

From the statistical analysis we find that coronal holes show a strong similarity in their DEM curves and derived properties. Thus, coronal holes can be can be clearly defined by their plasma properties. This not only enables a deeper understanding of the structure of coronal holes but can also serve to constrain the input for models \citep[e.g. solar wind models such as:][]{1999enlil,2018euhforia} and studies of coronal waves interacting with coronal holes \citep[e.g.][]{2019Podladchikova,2020piantschitsch}.


%
 \begin{acks}
 The SDO image data are available by courtesy of NASA and the respective science teams. S.G. Heinemann, M. Temmer and A.M. Veronig acknowledge funding by the Austrian Space Applications Programme of the Austrian Research Promotion Agency FFG (859729, SWAMI). J. Saqri acknowledges the support by the Austrian Science Fund FWF (I 4555).
 \end{acks}
%
\clearpage
\pagebreak
\newpage

 \appendix
The PSF and eclipse corrections could be introducing systematic biases, which might affect the correlations between the coronal hole plasma parameters with the other coronal hole parameters and the solar activity. To verify that we are not introducing such biases, we calculated the DEM using different input configurations. We used two different PSF kernels: firstly the PSF provided in the SSWIDL, which we used throughout the study, and secondly the PSF by \cite{2013Poduval}. We then performed the analysis using different values for the stray-light correction: We used the PSF-deconvoluted images without eclipse correction, with the eclipse correction as derived in Section~\ref{subs:psf}, and additionally with the method used by \citeauthor{2020saqri} (2020; with a correction factor of $[0.2,0.2,6.4,5.1,1.2,0.0]$ DN for the six SDO/AIA wavelengths respectively), who derived the correction from the 2012 Venus transit. The calculations were performed on a plate scale of \SI{9.6}{\arcsecond} (in contrast to the \SI{4.8}{\arcsecond} in the main study, which explains the small discrepancy to the values presented above) to complete the analysis in a reasonable amount of time. The resulting bootstrapped Pearson correlation coefficients for the average coronal hole temperature, electron density, and emission measure are presented in Tables~\ref{tab:cc_T},~\ref{tab:cc_n},~and~\ref{tab:cc_em} respectively. We find that the data preparation, i.e. which PSF was used and what correction is applied, does not significantly change the correlations, and as such the results are reliable.

\begin{table}
\caption{Pearson correlation coefficients for the average coronal hole temperature against coronal hole area, signed mean magnetic field density, and solar activity approximated by the international sunspot number and calculated from different input configurations. The correlation coefficients are given as calculated with and without uncertainties. In the square brackets the CI$_{90\%}$ are given.}\label{tab:cc_T}
\begin{threeparttable}
\begin{tabular}{l r r r} 
\multicolumn{4}{c}{Pearson Correlation Coefficients with Uncertainties} \\
 Correction & T vs. A & T vs. $\mid$B$_{\mathrm{s}}\mid$ & T vs. SSNr \\ \hline
\textsc{aia psf} & $-0.05~[-0.11,0.01]$ & $0.12~[0.06,0.01]$ & $0.16~[0.09,0.01]$ \\
\textsc{aia psf} + corr\tnote{a} & $-0.04~[-0.10,0.02]$ & $0.10~[0.04,0.02]$ & $0.13~[0.07,0.02]$ \\
\textsc{aia psf} + \citeauthor{2020saqri} -- corr\tnote{b} & $-0.05~[-0.11,0.01]$ & $0.16~[0.10,0.01]$ & $0.16~[0.10,0.01]$ \\
\citeauthor{2013Poduval} PSF & $-0.04~[-0.11,-0.02]$ & $0.12~[0.06,-0.02]$ & $0.15~[0.09,-0.02]$ \\
\citeauthor{2013Poduval} PSF + corr\tnote{a} & $-0.03~[-0.10,-0.01]$ & $0.10~[0.04,-0.01]$ & $0.14~[0.07,-0.01]$ \\

\hline \hline
\multicolumn{4}{c}{Pearson Correlation Coefficients without Uncertainties} \\
 Correction & T vs. A & T vs. $\mid$B$_{\mathrm{s}}\mid$ & T vs. SSNr \\ \hline
\textsc{aia psf} & $-0.15~[-0.19,-0.11]$ & $0.35~[0.25,-0.11]$ & $0.51~[0.45,-0.11]$ \\
\textsc{aia psf} + corr\tnote{a} & $-0.14~[-0.18,-0.09]$ & $0.34~[0.25,-0.09]$ & $0.52~[0.46,-0.09]$ \\
\textsc{aia psf} + \citeauthor{2020saqri} -- corr\tnote{b} & $-0.12~[-0.17,-0.08]$ & $0.43~[0.35,-0.08]$ & $0.51~[0.47,-0.08]$ \\
\citeauthor{2013Poduval} PSF & $-0.10~[-0.15,-0.08]$ & $0.27~[0.19,-0.08]$ & $0.41~[0.34,-0.08]$ \\
\citeauthor{2013Poduval} PSF + corr\tnote{a} & $-0.09~[-0.14,-0.07]$ & $0.27~[0.18,-0.07]$ & $0.41~[0.34,-0.07]$ \\
\hline
\end{tabular}
\begin{tablenotes}
          \item[a] Stray-light correction derived from lunar eclipse data as given in Table~\ref{tab:psf}.
          \item[b] Stray-light correction derived from Venus transit data as given given by \cite{2020saqri}.
        \end{tablenotes}
\end{threeparttable}
\end{table}

\begin{table}
\caption{Pearson correlation coefficients for the coronal hole electron density, against coronal hole area, signed mean magnetic field density, and solar activity approximated by the international sunspot number and calculated from different input configurations. The correlation coefficients are given as calculated with and without uncertainties. In the square brackets the CI$_{90\%}$ are given.}\label{tab:cc_n}
\begin{threeparttable}
\begin{tabular}{l|c c c} 
\multicolumn{4}{c}{Pearson Correlation Coefficient with Uncertainties} \\
 Correction & $n_\mathrm{e}$ vs. A & $n_\mathrm{e}$ vs. $\mid$B$_{\mathrm{s}}\mid$ & $n_\mathrm{e}$ vs. SSNr \\ \hline
\textsc{aia psf} & $-0.01~[-0.08,0.05]$ & $-0.14~[-0.19,0.05]$ & $0.03~[-0.04,0.05]$ \\
\textsc{aia psf} + corr\tnote{a} & $-0.02~[-0.08,0.05]$ & $-0.12~[-0.18,0.05]$ & $0.04~[-0.02,0.05]$ \\
\textsc{aia psf} + \citeauthor{2020saqri} -- corr\tnote{b} & $-0.02~[-0.08,0.04]$ & $-0.12~[-0.18,0.04]$ & $0.06~[-0.00,0.04]$ \\
\citeauthor{2013Poduval} PSF & $-0.02~[-0.09,0.04]$ & $-0.11~[-0.16,0.04]$ & $0.01~[-0.05,0.04]$ \\
\citeauthor{2013Poduval} PSF + corr\tnote{a} & $-0.02~[-0.09,0.04]$ & $-0.11~[-0.16,0.04]$ & $0.03~[-0.03,0.04]$ \\

\hline \hline
\multicolumn{4}{c}{Pearson Correlation Coefficient without Uncertainties} \\
 Correction & $n_\mathrm{e}$ vs. A & $n_\mathrm{e}$ vs. $\mid$B$_{\mathrm{s}}\mid$ & $n_\mathrm{e}$ vs. SSNr \\ \hline
\textsc{aia psf} & $-0.03~[-0.08,0.03]$ & $-0.28~[-0.33,0.03]$ & $0.07~[0.00,0.03]$ \\
\textsc{aia psf} + corr\tnote{a} & $-0.04~[-0.09,0.02]$ & $-0.26~[-0.31,0.02]$ & $0.11~[0.05,0.02]$ \\
\textsc{aia psf} + \citeauthor{2020saqri} -- corr\tnote{b} & $-0.04~[-0.10,0.01]$ & $-0.25~[-0.30,0.01]$ & $0.15~[0.08,0.01]$ \\
\citeauthor{2013Poduval} PSF & $-0.03~[-0.10,0.03]$ & $-0.16~[-0.21,0.03]$ & $0.02~[-0.05,0.03]$ \\
\citeauthor{2013Poduval} PSF + corr\tnote{a} & $-0.04~[-0.10,0.03]$ & $-0.17~[-0.23,0.03]$ & $0.06~[-0.01,0.03]$ \\
 
\hline
\end{tabular}
\begin{tablenotes}
          \item[a] Stray-light correction derived from lunar eclipse data as given in Table~\ref{tab:psf}.
          \item[b] Stray-light correction derived from Venus transit data as given given by \cite{2020saqri}.
        \end{tablenotes}
\end{threeparttable}
\end{table}

\begin{table}
\caption{Pearson correlation coefficients for the coronal hole emission measure against coronal hole area, signed mean magnetic field density, and solar activity approximated by the international sunspot number, and calculated from different input configurations. The correlation coefficients are given as calculated with and without uncertainties. In the square brackets the CI$_{90\%}$ are given.}\label{tab:cc_em}
\begin{threeparttable}
\begin{tabular}{l|c c c} 
\multicolumn{4}{c}{Pearson Correlation Coefficient with Uncertainties} \\
 Correction & EM vs. A & EM vs. $\mid$B$_{\mathrm{s}}\mid$ & EM vs. SSNr \\ \hline
\textsc{aia psf} & $-0.02~[-0.08,0.04]$ & $-0.09~[-0.15,0.04]$ & $0.09~[0.03,0.04]$ \\
\textsc{aia psf} + corr\tnote{a} & $-0.01~[-0.08,0.05]$ & $-0.09~[-0.15,0.05]$ & $0.09~[0.02,0.05]$ \\
\textsc{aia psf} + \citeauthor{2020saqri} -- corr\tnote{b} & $-0.02~[-0.08,0.05]$ & $-0.08~[-0.14,0.05]$ & $0.10~[0.04,0.05]$ \\
\citeauthor{2013Poduval} PSF & $-0.02~[-0.08,0.05]$ & $-0.09~[-0.15,0.05]$ & $0.08~[0.02,0.05]$ \\
\citeauthor{2013Poduval} PSF + corr\tnote{a} & $-0.01~[-0.07,0.05]$ & $-0.10~[-0.15,0.05]$ & $0.08~[0.02,0.05]$ \\

\hline \hline
\multicolumn{4}{c}{Pearson Correlation Coefficient without Uncertainties} \\
 Correction & EM vs. A & EM vs. $\mid$B$_{\mathrm{s}}\mid$ & EM vs. SSNr \\ \hline
\textsc{aia psf} & $-0.04~[-0.09,0.02]$ & $-0.17~[-0.23,0.02]$ & $0.21~[0.15,0.02]$ \\
\textsc{aia psf} + corr\tnote{a} & $-0.03~[-0.08,0.03]$ & $-0.19~[-0.25,0.03]$ & $0.21~[0.15,0.03]$ \\
\textsc{aia psf} + \citeauthor{2020saqri} -- corr\tnote{b} & $-0.03~[-0.09,0.02]$ & $-0.17~[-0.22,0.02]$ & $0.25~[0.20,0.02]$ \\
\citeauthor{2013Poduval} PSF & $-0.03~[-0.09,0.03]$ & $-0.15~[-0.20,0.03]$ & $0.16~[0.10,0.03]$ \\
\citeauthor{2013Poduval} PSF + corr\tnote{a} & $-0.02~[-0.08,0.04]$ & $-0.18~[-0.23,0.04]$ & $0.18~[0.11,0.04]$ \\
\hline
\end{tabular}
\begin{tablenotes}
          \item[a] Stray-light correction derived from lunar eclipse data as given in Table~\ref{tab:psf}.
          \item[b] Stray-light correction derived from Venus transit data as given given by \cite{2020saqri}.
        \end{tablenotes}
\end{threeparttable}
\end{table}

\footnotesize\paragraph*{Disclosure of Potential Conflicts of Interest}
The authors declare that they have no conflicts of interest.

%
%
%
%
%
%

\end{article} 
\end{document}